\documentclass[twocolumn,showpacs,preprintnumbers,amsmath,amssymb]{revtex4}

\usepackage{graphicx}
\usepackage{dcolumn}
\usepackage{bm}

\begin{document}


\title{Cell dynamics approach to the formation of metastable phases 
during phase transformation }

\author{M. Iwamatsu}
\affiliation{
Department of Physics, General Education Center,
Musashi Institute of Technology,
Setagaya-ku, Tokyo 158-8557, Japan
}%


\date{\today}

\begin{abstract}
In this paper, we use the cell dynamics method to study the dynamics of 
phase transformation when three phases exist.  The system we study is 
a two-dimensional system.  The system is able to achieve three phases 
coexistence, which for simplicity we call crystal, liquid and vapor phases.  
We focus our study on the case when the vapor and crystal phases are stable 
and can coexist while the other intermediate liquid phase is metastable.  
In this study we examine the most fundamental process of the growth of 
a composite nucleus which consists of a circular core of one phase 
surrounded by a circular layer of second phase embedded in a third phase.  
We found that there is one special configuration that consists of a core 
stable phase surrounded by another stable phase in a metastable liquid 
environment which becomes stationary and stable.  Then, the nucleus does not 
grow and the metastable liquid survives.  The macroscopic liquid phase does 
not disappear even though it is thermodynamically metastable.  This result 
seems compatible to the argument of kinetics of phase transition developed 
by Cahn [J. Am. Ceram. Soc. {\bf 52}, 118 (1969)] based on the construction 
of a common tangent of the free energy curve.
\end{abstract}

\maketitle

\section{\label{sec:sec1}Introduction}
\label{sec:level1}

The study of the formation of the long-lived metastable phase during phase 
transformation has a long history~\cite{Ostwald,Cahn}.  Recently, renewed 
interest in the metastable phase formation, in particular, in the field 
of soft-condensed matter physics~\cite{Poon, Evans1} has emerged because of 
the rather long relaxation time of these materials, which, typically, are 
in the ranges  1 ms to 1 year~\cite{Renth}.  Although the formation of the 
thermodynamically metastable phase is not only academically but 
industrially important because many industrial products are in long-lived 
metastable state, the theoretical study of the kinetics of phase 
transformation is hindered because of the lack of appropriate theoretical 
and computational models.  Hence, a detailed understanding of the kinetics 
of phase transformation using realistic modeling is essential. 

The direct microscopic computer simulation of the nucleation and the kinetics 
of phase transformation using molecular dynamics or the Monte Carlo method 
is possible~\cite{Auer} but is still a difficult task.  Even the most fundamental 
phenomenon like nucleation is still not easy. In order to avoid the demand for 
huge computational resources, and to get the qualitative (course-grained) 
picture of the kinetics of phase transformation, the mesoscopic approach 
based on the phenomenological model called the Cahn-Hilliard~\cite{Cahn2}, 
Ginzburg-Landau~\cite{Valls} or phase-field model~\cite{Castro,Granasy2}, 
which requires a solution using a non-linear partial differential equation, 
has been traditionally employed.  Since this approach requires the time integration 
of highly non-linear partial differential equations, it is still not easy 
to simulate the long-time behavior of the kinetics of phase 
transformation~\cite{Roger} except for the various forms of special analytical 
traveling wave solutions~\cite{Chan,Bechhoefer,Celestini,Iwamatsu}.

In order to understand the full kinetics of phase transformation with the transient 
and long-lived metastable phase, an efficient simulation method is absolutely 
necessary.  In this report, we use a formalism which is based on the cell dynamics 
method to investigate the kinetics of the metastable phase during the phase 
transformation when the circular grain (nucleus) of the stable phase grows.  
Our result suggests that the metastable phase can be long-lived indeed, and it 
also indicates that the cell dynamics method is efficient and flexible enough 
to study the kinetics of the metastable phase during the phase transformation.

\section{Cell dynamics method for three-phase system}
\label{sec:sec2}

In order to study the phase transformation, it is customary to study the 
partial differential equation called the time-dependent Ginzburg-Landau 
(TDGL) equation:
\begin{equation}
\frac{\partial \psi}{\partial t}=-\frac{\delta \mathcal{F}}{\delta \psi}
\label{eq:2-1}
\end{equation}
where $\psi$ is the {\it non-conserved} order parameter and $\mathcal{F}$ is the free energy 
functional (grand potential), which is usually written as the square-gradient form:
\begin{equation}
\mathcal{F}[\psi]=\int d{\bf r} \left[\frac{1}{2}D(\nabla \psi)^{2}+h(\psi)\right]
\label{eq:2-2}
\end{equation}
The local part of the free energy $h(\psi)$ determines the bulk phase diagram and 
the value of the order parameters in equilibrium phases. Traditionally, 
the double-well form
\begin{equation}
h(\psi) = -\frac{1}{2}\psi^2+\frac{1}{4}\psi^4
\label{eq:2-3}
\end{equation}
has been used to model the phase transformation of a two phase system.

Puri and Oono~\cite{Puri} transformed this TDGL equation (\ref{eq:2-1}) for the 
non-conserved order parameter into the space-time discretized cell-dynamics 
equation following a similar transformation of the kinetic equation for the 
conserved order parameter called the Cahn-Hilliard-Cook equation~\cite{Oono}.  
Their transformation does not correspond to the numerical approximation of 
the original TDGL equation.  Rather, they aimed at simulating the kinetics 
of phase transformation of real system within the framework of discrete 
cellular automata.  

According to their cell dynamics method, the partial differential equation 
(\ref{eq:2-1}) is transformed into the finite difference equation in space 
and time:
\begin{equation}
\psi(t+1,n)=F[\psi(t,n)]
\label{eq:2-4}
\end{equation}
where the time $t$ is discrete integer and the space is also discrete and 
is expressed by the site index (integer) $n$.  The mapping $F$ is given by
\begin{equation}
F[\psi(t,n)]=-f(\psi(t,n))+\left[<<\psi(t,n)>>-\psi(t,n)\right]
\label{eq:2-5}
\end{equation}
where $f(\psi)=dh(\psi)/d\psi$ and the definition of $<<*>>$ for the two-dimensional 
square grid is given by
\begin{equation}
<<\psi(t,n)>>=\frac{1}{6}\sum_{i=\mbox{nn}}\psi(t,i)
+\frac{1}{12}\sum_{i=\mbox{nnn}}\psi(t,i)
\label{eq:2-6}
\end{equation}
with "nn" means the nearest neighbors and "nnn" the next-nearest neighbors 
of the square grid.  Improved forms of this mapping function $F$ for 
three-dimensional case was also obtained~\cite{Puri,Oono,Teixeira1}.

Oono and Puri~\cite{Puri,Oono} have further approximated the derivative of 
the local free energy $f(\psi)$ called "map function" by the $\tanh$ form:
\begin{equation}
f(\psi)=\frac{dh}{d\psi}\simeq \psi-A \tanh \psi
\label{eq:2-7}
\end{equation}
with $A=1.3$, which corresponds to the free energy~\cite{Chakrabarti}:
\begin{equation}
h(\psi)=-A\ln\left(\cosh\psi\right)+\frac{1}{2}\psi^{2}
\label{eq:2-8}
\end{equation}
and is the approximation to (\ref{eq:2-3}) if $A=1.5$~\cite{Chakrabarti,Teixeira1}.  
Later Chakrabarti and Brown~\cite{Chakrabarti} argued that this simplification 
is justifiable since the detailed form (\ref{eq:2-3}) of the free energy $h(\psi)$ 
is irrelevant to the long-time kinetics and the scaling exponent.

Subsequently, however, several authors used the map function $f(\psi)$ directly 
obtained from the free energy $h(\psi)$ in cell dynamics equation (\ref{eq:2-5}) 
as it is~\cite{Qi,Ren1} and found that the cell dynamics equation is still 
amenable for a realistic map function numerically.   Ren and Hamley~\cite{Ren1} 
argued that by using the original form of the free energy function $f(\psi)$ one 
can easily include the effect of asymmetry of free energy and, hence, the 
asymmetric character of two phases can be considered.  It is now well recognized 
that this cell dynamics method can reproduce the essential feature of the 
kinetics of phase transformation between two phases even though the method 
is not guaranteed~\cite{Teixeira1} to be an accurate approximation of the 
original TDGL partial differential equation (\ref{eq:2-1}).

\begin{figure}[htbp]
\begin{center}
\includegraphics[width=0.7\linewidth]{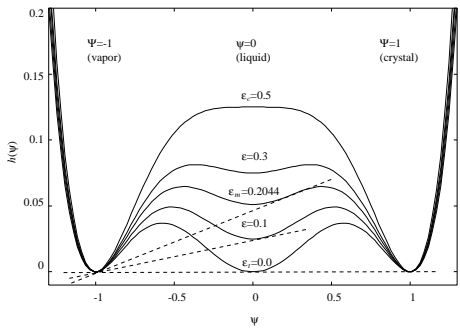}
\caption{
The model triple-well free energy~\cite{Widom} which can achieve a three-phase 
coexistence.  The left phase with $\psi_{v}=-1$ is called vapor, the central 
phase with $\psi_{l}=0$ is called liquid and the right phase with $\psi_{c}=1$ is 
called crystal for convenience.  The triple point is at $\epsilon_{t}=0$, when all 
three phases are at equilibrium and can coexist.  The liquid phase becomes 
metastable 
when $\epsilon_{t} < \epsilon < \epsilon_{m}$ where $\epsilon_{m}=0.2044$ 
though the local vapor-liquid or 
liquid-crystal equilibrium could be realized because the common tangent 
between the stable vapor and the metastable liquid phases or those between 
stable crystal and metastable liquid phases can be drawn 
(broken lines).  When the parameter $\epsilon$ exceeds $\epsilon_{m}=0.2044$, 
no common tangent can be drawn, so the two-phase 
equilibrium between vapor-liquid and liquid-crystal cannot be achieved.  
When $\epsilon$ is larger than $\epsilon_{c}=0.5$, the liquid phase becomes completely 
unstable.
}
\label{Fig:1}
\end{center}
\end{figure}

A further extension of the cell dynamics equation to the three phase system is 
simple.  One has to introduce the free energy function $h(\psi)$ of triple-well 
form, which can achieve a three-phase coexistence.  In our report, we will use 
one of the simplest analytical forms proposed by Widom~\cite{Widom}:
\begin{equation}
h(\psi) = \frac{1}{4}(\psi+1)^{2}(\psi-1)^{2}(\psi^{2}+\epsilon)
\label{eq:2-8-0}
\end{equation}
where the parameter $\epsilon$ controls the relative stability of three phases.  
Several shapes of the free energy function $h(\psi)$ for several values of 
the parameter $\epsilon$ are shown in Fig.~\ref{Fig:1}.  As can be seen from 
the figure, two phases around $\psi_{v}=-1$ which we call {\it vapor} and $\psi_{c}=1$ 
which we call {\it crystal} for simplicity always coexist, while another phase 
around $\psi_{l}=0$ which we call {\it liquid} can be metastable when 
$0\leq \epsilon \leq 0.5$.  
The free energy of this metastable liquid phase is higher than that of the 
stable crystal or vapor phases by the amount
\begin{equation}
\Delta h = f(\psi=1) - f(\psi=0) = \frac{\epsilon}{4}
\label{eq:2-8-1}
\end{equation}

This liquid phase becomes completely unstable and disappears when $\epsilon>\epsilon_{c}$
where $\epsilon_{c}=0.5$ is the critical point.  
When $\epsilon<0$, only the liquid phase is stable and both the vapor and the 
crystal phases are metastable instead.  Since we are interested in the case 
when only one intermediate phase is metastable, we will consider the case 
when $0\leq \epsilon \leq \epsilon_{c}$.  As will be shown in the next section, even 
though the metastable liquid minimum is irrelevant for the equilibrium phase 
behavior, it not only controls the phase transition kinetics but appears as 
the long-lived macroscopic metastable phase during the phase transformation.

Similar triple-well potentials were used by several workers to study the 
nucleation~\cite{Granasy} and the metastable phase formation~\cite{Evans1,Celestini,Evans4} 
within the framework of the original TDGL or the phase-field model.  The 
importance of this triple-well free energy and the appearance of the 
metastable state were also recently suggested experimentally in a 
colloid-polymer mixture~\cite{Poon,Renth}

\section{Numerical results and discussions}
\label{sec:sec3}

\subsection{Front velocity of the growing stable phase}

Before looking at the issue of the kinetics of phase transformation of 
a metastable phase in a three-phase system, we will briefly look at 
the growth of one circular nucleus from the stable phase after nucleation 
in a two-phase system. In order to simulate the evolution of a nucleus, we 
have to prepare the system as a two-phase system in which one phase is stable and another 
is metastable and has higher free energy than the former.  The free 
energy difference between the stable and metastable phases is controlled 
by the super saturation in usual liquid condensation from vapor and by 
the under-cooling in usual crystal nucleation.  Microscopically, this 
free energy difference is necessary for the nucleus of the stable phase 
to grow by overcoming the curvature effect of the surface tension~\cite{Iwamatsu2}.

\begin{figure}[htbp]
\begin{center}
\includegraphics[width=0.7\linewidth]{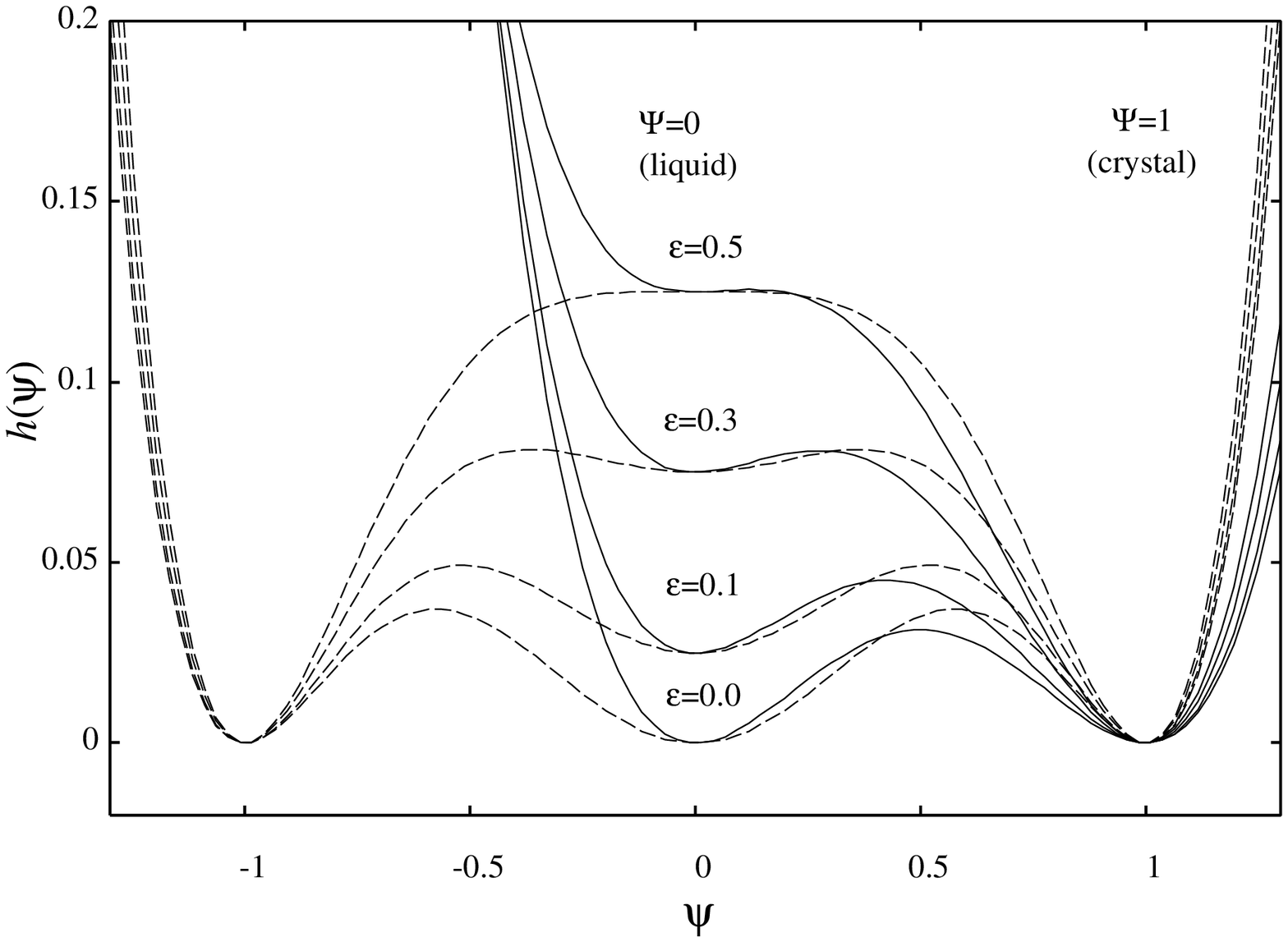}
\caption{
The model double-well free energy~\cite{Jou,Iwamatsu3} (solid curve) which 
achieves the two-phase equilibrium is compared with the triple-well free 
energy in Fig.~\ref{Fig:1} (broken curve).  The parameter $\epsilon$ determines 
the free energy difference $\Delta h$, and the parameter $q$ determines 
the free energy barrier. We choose $q=2$ to mimic the triple-well potential.
}
\label{Fig:2-1}
\end{center}
\end{figure}

In order to study the growth of a stable phase using a cell dynamics system, we 
will consider the time-dependent Ginzburg-Landau (TDGL) equation (\ref{eq:2-1}) 
of square gradient form (\ref{eq:2-2}).  The local part of the free energy $h(\psi)$, 
which we use is~\cite{Jou,Iwamatsu3}:
\begin{equation}
h(\psi) = \frac{1}{4}q\psi^{2}(1-\psi)^{2} 
+ \frac{3}{2}\epsilon\left(\frac{\psi^{3}}{3}-\frac{\psi^{2}}{2}\right)
+\frac{1}{4}\epsilon
\label{eq:3-1}
\end{equation}
This free energy is shown in Fig.~\ref{Fig:2-1} where the liquid phase 
with $\psi_{l}=0$ is metastable while the crystal phase with $\psi_{c}=1$ is stable, 
which mimics the liquid-crystal part of the triple-well potential (\ref{eq:2-8-0}).  
The free energy barrier between two phases can be tuned by $q$, while the 
free energy difference $\Delta h$ between the stable phase at $\psi_{c}=1$ and the 
metastable phase at $\psi_{l}=0$ can be controlled by $\epsilon$.  This free 
energy difference is given by the same formula (\ref{eq:2-8-1}) as in the 
model three-phase system.  We choose the parameter $q=2$ in order to mimic 
the functional form of the triple well free energy in Fig.~\ref{Fig:1} as 
shown in Fig.~\ref{Fig:2-1}.

The time-dependent Ginzburg-Landau (TDGL) equations (\ref{eq:2-1}) and (\ref{eq:2-2}) for a circular or a spherical
growing nucleus of stable phase in a metastable environment with radial coordinate $r$ is written as~\cite{Chan}
\begin{equation}
D\left(\frac{\partial^{2}}{\partial r^{2}}+\frac{d-1}{r}\frac{\partial}{\partial r}\right)\psi-\frac{\partial\psi}{\partial t}
=\frac{\partial h}{\partial \psi}
\label{eq:a1}
\end{equation}
where $d$ is the dimension ($d=2$ for a circular and $d=3$ for a spherical nucleus) of the problem.  
Then, the traveling wave solution having radial symmetry with moving interface at $R(t)$ of the form
\begin{equation}
\psi({\bf r},t)=\psi(X), 
\label{eq:a2}
\end{equation}
with $X=r-R(t)$ satisfies the differential equation
\begin{equation}
D\frac{d^{2}\psi}{dX^{2}}+v\frac{d\psi}{dX}-\frac{\partial h}{\partial \psi}=0
\label{eq:a3}
\end{equation}
with
\begin{equation}
v = \frac{dR}{dt}+\frac{D(d-1)}{R}.
\label{eq:a4}
\end{equation}
Equation (\ref{eq:a3}) represents a mechanical analogue of the equation of
motion of classical particle in a potential well $-h$ subject to a friction force
which is proportional to the parameter $v$.  Therefore, a finite size of the free energy
difference $\Delta h$ is necessary for (\ref{eq:a3}) to compensate for the dissipation
of energy due to friction and to have a solution which corresponds to a traveling wave.  In other words,
the moving (growing or shrinking) interface is possible only when
there is the free energy difference $\Delta h$ between two phases.  Therefore, one phase should be metastable and another should be stable.

Equation (\ref{eq:a3}) has a particular solution only when the 
parameter $v$ takes a specific value.
The corresponding interfacial velocity $dR/dt$ is given by
\begin{equation}
\frac{dR}{dt}=v-\frac{D(d-1)}{R}
\label{eq:a5}
\end{equation}
where, the second term 
on the right hand side represents the effect of capillary pressure.
For a
larger nucleus with $R\rightarrow \infty$, the interfacial velocity becomes
$v$ which is also the formula for the one-dimension problem with $d=1$.
For a smaller nucleus, the actual interfacial velocity $dR/dt$ will be smaller than $v$ due to the capillary pressure because $R>0$.  In particular, 
when $dR/dt<0$, the nucleus cannot grow.  The nucleus with a radius
$R$ larger than the
critical radius $R_{c}$ grows while the one with a smaller radius disappears.
The critical radius $R_{c}$ is determined from $dR/dt=0$ which gives
\begin{equation}
R_{c}=\frac{D(d-1)}{v}
\label{eq:a6}
\end{equation}    
Therefore, in two-phase coexistence with $v=0$, any circular or spherical nucleus 
with finite radius $R$ disappears, and only a flat interface remains. 

The shrinking metastable void within a stable phase is also described
by the equations (\ref{eq:a1}) to (\ref{eq:a6}), but now, with $v<0$ and $dR/dt<0$.
Then, the capillary pressure in (\ref{eq:a5}) always accelerates the
interfacial velocity, and there is no critical radius $R_{c}$ for the void.  

The above steady-state solution of TDGL with a constant interfacial velocity $v$ was 
obtained analytically in a one-dimension by Chan~\cite{Chan} when the free 
energy is written using the quartic form (\ref{eq:3-1}).   Using his formula, 
the interfacial velocity $v$ of our TDGL model (\ref{eq:2-1}) and (\ref{eq:2-2}) 
with the free energy (\ref{eq:3-1}) is given by:
\begin{equation}
v = \sqrt{\frac{D}{2q}}3\epsilon
\label{eq:3-2}
\end{equation}    
Chan~\cite{Chan} further suggested that if the interfacial width is narrow, 
the interfacial velocity of a circular or spherical growing nucleus is 
asymptotically given by the same formula (\ref{eq:3-2}).  The larger the 
free energy difference $\epsilon$, and the lower the free energy 
barrier $q$, the higher the front velocity $v$ from (\ref{eq:3-2}).

The critical radius $R_{c}$ of a circular nucleus in a two-dimensional system 
is also given  analytically~\cite{Chan,Jou} by
\begin{equation}
R_{c}=\frac{D}{v}=\frac{\sqrt{2qD}}{3\epsilon}
\label{eq:3-3}
\end{equation}
In the metastable environment, a nucleus of a stable phase with a radius $R$ 
smaller than $R_{c}$ shrinks, while the nucleus with a radius larger 
than $R_{c}$ grows and its front velocity approaches (\ref{eq:3-2}).  
The void of a metastable phase surrounded by a stable environment always 
shrinks regardless of the size of the critical radius.  Again, the larger 
the free energy difference $\epsilon$, and the lower the free energy 
barrier $q$, the smaller the critical radius $R_{c}$ . 

We have imported the above free energy (\ref{eq:3-1}) into the cell dynamics 
code written by Mathematica TM~\cite{Wolfram} for the animation of spinodal 
decomposition developed by Gaylord and Nishidate~\cite{Gaylord}, and simulated 
the growth of the stable crystal phase in a metastable liquid environment and 
that of a metastable liquid void in a stable crystal environment.

\begin{figure}[htbp]
\begin{center}
\includegraphics[width=0.7\linewidth]{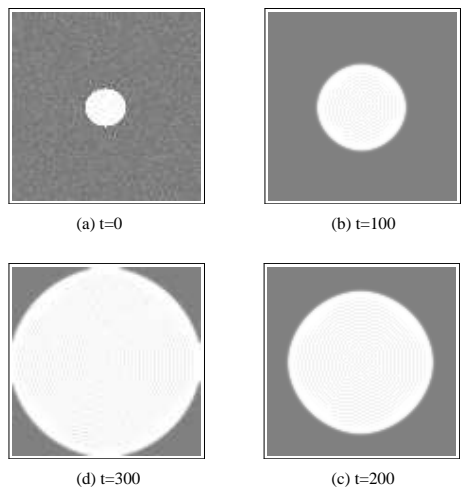}
\caption{
The evolution of the crystal nucleus (white) in the metastable liquid 
environment (gray) when $q=2$ and $\epsilon=0.1$.}
\label{Fig:2-2}
\end{center}
\end{figure}

\begin{figure}[htbp]
\begin{center}
\includegraphics[width=0.7\linewidth]{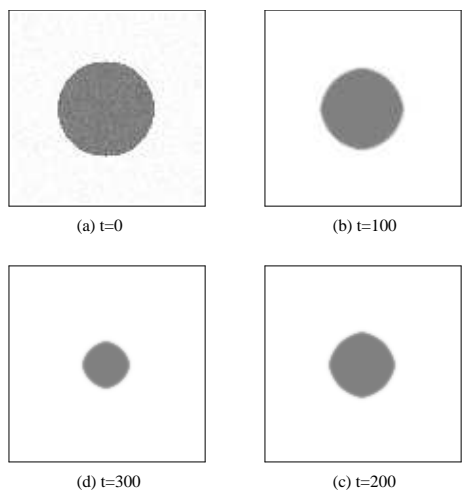}
\caption{
The contraction of the metastable liquid void (gray) in the stable crystal 
environment (white) when $q=2$ and $\epsilon=0.1$. }
\label{Fig:2-2x}
\end{center}
\end{figure}

Figures~\ref{Fig:2-2} and \ref{Fig:2-2x} show the evolution of the stable 
circular crystal phase in a metastable liquid environment, and the 
contraction of a metastable liquid void in a stable crystal environment. 
The system size is 100$\times$100=10000 and $D=0.5$~\cite{Puri,Gaylord}.
The periodic boundary condition is used.
The initial nucleus or void is prepared by randomly selecting the order 
parameter $\psi$ from $0.9\leq \psi \leq 1.1$ for stable crystal and 
from $-0.1\leq \psi \leq 0.1$ for metastable liquid.  The initial random 
distribution is necessary because our cell-dynamics system is deterministic 
and does not include random noise.  The figures~\ref{Fig:2-2} and \ref{Fig:2-2x} 
show that the stable circular crystal grows and the metastable circular 
liquid shrinks steadily without changing the circular shape appreciably.  

\begin{figure}[htbp]
\begin{center}
\includegraphics[width=0.7\linewidth]{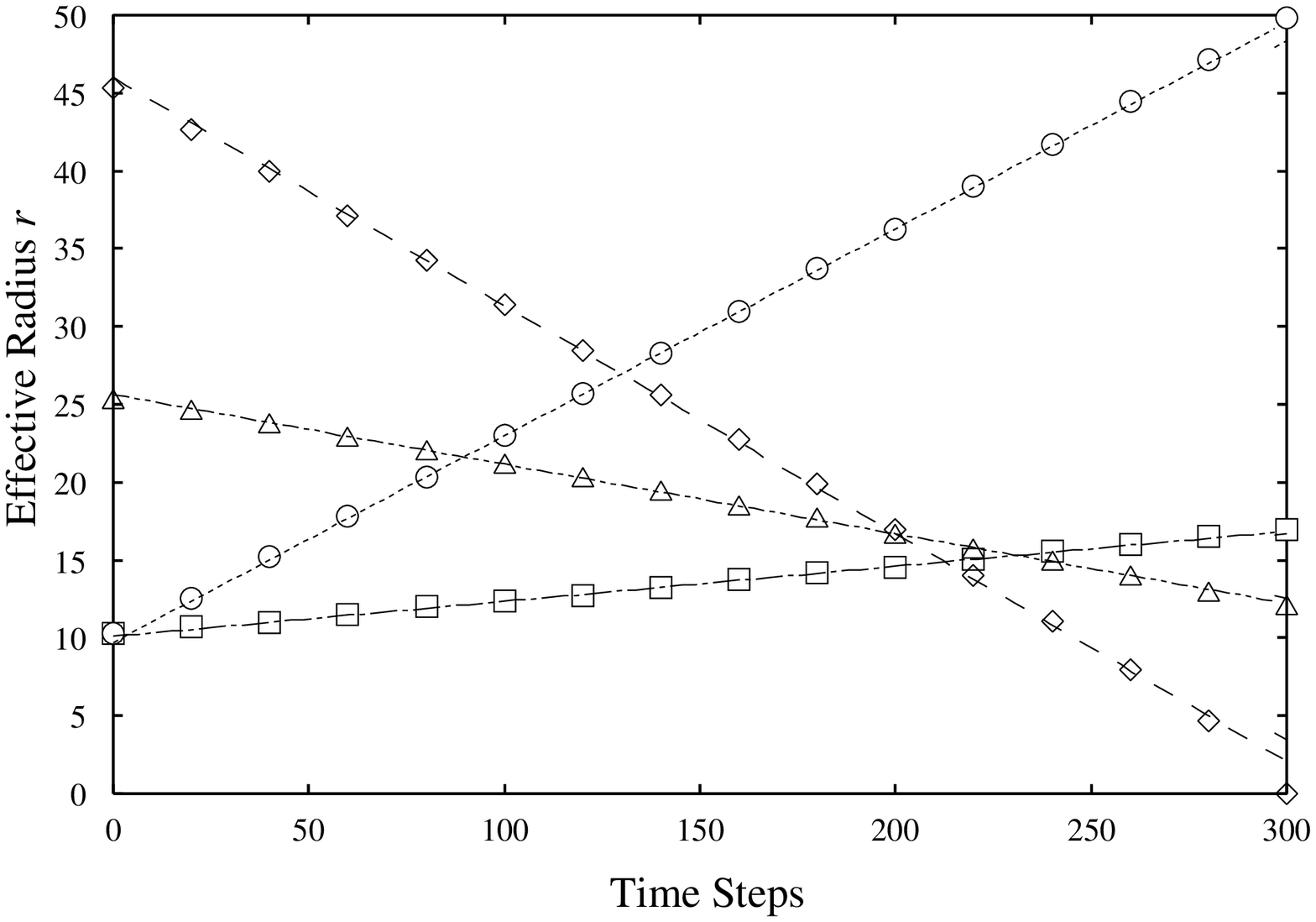}
\caption{
The evolution and contraction of the effective radii of the crystal nuclei 
and liquid voids calculated from (\ref{eq:3-3x}) plotted as the function of 
time steps. $\square$: crystal nucleus when $\epsilon=0.1$, $\bigcirc$: crystal 
nucleus when $\epsilon=0.3$, $\triangle$: liquid void when $\epsilon=0.1$, 
$\lozenge$: liquid void when $\epsilon=0.3$. $q$ is set to $q=2$ for all 
four cases.  The total area is 100$\times$100=10000 pixels.  The straight 
lines are the least-square fittings to the numerical data. 
}
\label{Fig:2-3}
\end{center}
\end{figure}

In Fig.~\ref{Fig:2-3} the effective radius $r$ of the circular nucleus of a 
stable crystal grain and a metastable liquid void, which is calculated by 
assuming the circular area $S$ from
\begin{equation}
r = \sqrt{\frac{S}{\pi}}
\label{eq:3-3x}
\end{equation}
is plotted as the function of the time step.  We defined the area $S$ of 
crystal as the number of pixels whose order parameter $\psi$ is larger 
then 0.5.  Figure~\ref{Fig:2-3} clearly indicates a nearly linear growth 
of the radius $r$ of the stable phase which means the constant front velocity 
of the liquid-crystal interface.   

The velocities $v$ estimated from Fig.~\ref{Fig:2-3} are summarized in 
Tab.~\ref{tab:1}. Table~\ref{tab:1} shows that the analytical expression 
in eq.(\ref{eq:3-2}) gives a rough estimate of the front velocity. The velocity
of shrinking void is always larger than the growing nucleus due to the
capillary pressure as expected. 
Understanding that the cell dynamics method {\it does not} attempt to 
solve the original TDGL directly and, hence, is not guaranteed to 
reproduce the analytical expression (\ref{eq:3-2}), the discrepancy 
between the cell dynamics simulation and theoretical prediction 
in (\ref{eq:3-2}) seems not so serious.  There is also a problem of the 
definition of the area $S$ of the growing phase, which will also 
numerically affect the front velocity calculated from (\ref{eq:3-3x}).

\begin{table}[htb]
\begin{center}
\caption{
The front velocities of the growing new stable phase (nucleus) and shrinking 
unstable phase (void) obtained from the cell dynamics simulation of 
Fig.~\ref{Fig:2-3} compared with the theoretical prediction from eq.(\ref{eq:3-2}).  
The critical radii calculated from eq.(\ref{eq:3-3}) are also shown for reference. }
\label{tab:1}
\begin{tabular}{ccccc}
\hline
$\epsilon$ & $R_{c}$ & $v$ & $v_{\rm crystal-liquid}$ & $v_{\rm crystal-liquid}$ \\
      & (Theoretical) & (Theoretical) & (growing) & (shrinking) \\
\hline
0.1 & 4.71 & 0.106 & 0.023 & 0.045 \\
0.3 & 1.57 & 0.318 & 0.133 & 0.146 \\
\hline
\end{tabular}
\end{center}
\end{table}

From the comparison of the cell dynamics simulation and the theoretical prediction 
from the TDGL for the two-phase system, we have confidence that this cell dynamics 
method should be effective to study the qualitative features of the evolution of 
the metastable phase in a three-phase system, which we will discuss in the next 
subsection of this paper.  More details about the application of the cell dynamics 
method to the phase transformation in a two-phase system and the simulation of 
the so-called Kolmogorov-Johnson-Mehl-Avram (KJMA) kinetic will be presented 
elsewhere~\cite{Iwamatsu3}.

\subsection{Long-lived metastable phase in a three-phase system}

Since we are most interested in the evolution or regression of the metastable 
phase during the phase transformation after nucleation, we will study the kinetics 
of phase transformation when three phases exist using the model free energy 
defined by eq.(\ref{eq:2-8-0}) and depicted in Fig.~\ref{Fig:1}.

\begin{figure}[htbp]
\begin{center}
\includegraphics[width=0.7\linewidth]{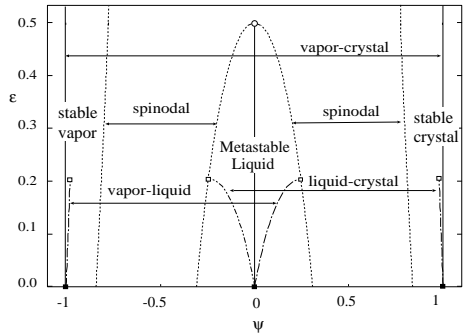}
\caption{
The phase diagram for the model triple-well free energy defined by (\ref{eq:2-8-0}).  
The model realizes the stable vapor and crystal phase at $\psi_{v}=-1$ (solid vertical line) 
and $\psi_{c}=1$ (solid vertical line) 
respectively and the metastable liquid phase at $\psi_{l}=0$ (solid vertical line) 
if $0<\epsilon<\epsilon_{c}$ with $\epsilon_{c}=0.5$.  
The triple point is at $\epsilon_{t}=0$ when three phases ($\blacksquare$) are all stable 
and can coexist.  The metastable liquid ceases to exist at $\psi=0$ and 
$\epsilon_{c}=0.5$ which is the hidden critical point ($\bigcirc$).  There are not 
only the vapor-crystal binodal (two vertical solid lines) but the hidden vapor-(metastable) 
liquid binodal (single-dot chain curve) 
and hidden (metastable)liquid-crystal binodal (double-dots chain curve), 
which can be constructed from 
the common tangent as shown in Fig.~\ref{Fig:1}.  These two pars of binodal disappear ($\square$) when
they terminate the spinodal lines (dotted curves) at $\epsilon_{m}=0.2044$. The spinodal regions are also 
complex because of the existence of the metastable liquid phase.
}
\label{Fig:2}
\end{center}
\end{figure}

The phase diagram of this system which we will study is shown in Fig.~\ref{Fig:2}.  
The equilibrium stable vapor phase has an order parameter $\psi_{v}=-1$ and the 
stable crystal phase has $\psi_{c}=1$, then the vapor-crystal coexistence lines 
(binodal) are the vertical lines at $\psi_{v}=-1$ and $\psi_{c}=1$.  However, the stable liquid 
at $\psi_{l}=0$ exists only at the triple point $\epsilon_{t}=0$.

Once the stable liquid phase disappears from the equilibrium phase diagram 
when $\epsilon>0$, it can be considered to be {\it hidden} or {\it buried} as a 
metastable liquid phase within the vapor-crystal binodal. Even this 
metastable liquid cannot exist if $\epsilon>\epsilon_{c}$ where $\epsilon_{c}=0.5$. 
The hidden critical 
point is at $\epsilon_{c}=0.5$ and $\psi=0$, where the {\it metastable} 
liquid becomes completely {\it unstable}.  

There are also hidden 
vapor-liquid and a liquid-crystal local coexistence lines (binodals), 
which are given by the co-tangency points $\psi_{v}^{'}$ and $\psi_{l}^{'}$ 
of the common tangent between 
the vapor and liquid free energy and by $\psi_{l}^{''}$ and $\psi_{c}^{''}$ 
between the liquid and crystals 
energy respectively.  These hidden binodals disappear at 
$\epsilon=\epsilon_{m}$ with $\epsilon_{m}=0.2044$ where 
the solutions of the simultaneous equations, for example
\begin{eqnarray}
h^{'}\left(\psi_{l}^{''}\right)&=&h^{'}\left(\psi_{c}^{''}\right)
\nonumber \\
h\left(\psi_{l}^{''}\right)-h^{'}\left(\psi_{l}^{''}\right)\psi_{l}^{''} &=& 
h\left(\psi_{c}^{''}\right)-h^{'}\left(\psi_{c}^{''}\right)\psi_{c}^{''}
\label{eq:b1x}
\end{eqnarray}
ceases to exist, and the hidden binodal lines are 
terminated by the spinodal lines as shown in Fig.~\ref{Fig:2}.  Then, the vapor-liquid and the 
liquid-crystal coexistence could be established locally even when 
the liquid phase is metastable so long as $\epsilon<\epsilon_{m}$.

For the free energy (\ref{eq:2-8}), the spinodal lines are defined by the condition:
\begin{equation}
\frac{d^{2}h}{d\psi^{2}}=0
\label{eq:2-9}
\end{equation}
which gives the outer spinodal lines $\psi_{\rm sp1}$:
\begin{equation}
\psi_{\rm sp1} = \pm \sqrt{\frac{2}{5}-\frac{\epsilon}{5}+\frac{\sqrt{7-2\epsilon+3\epsilon^{2}}}{5\sqrt{3}}}
\label{eq:2-10}
\end{equation}
and inner spinodal lines $\psi_{\rm sp2}$:
\begin{equation}
\psi_{\rm sp2} = \pm \sqrt{\frac{2}{5}-\frac{\epsilon}{5}-\frac{\sqrt{7-2\epsilon+3\epsilon^{2}}}{5\sqrt{3}}}
\label{eq:2-10-1}
\end{equation}
The latter merge at $\psi_{l}=0$ and $\epsilon_{c}=0.5$.  Therefore, the spinodal region 
consists of two regions sandwiched by an outer spinodal and an inner spinodal 
lines when $\epsilon<\epsilon_{c}$ while it consists of one region for $\epsilon>\epsilon_{c}$ as 
shown in Fig.~\ref{Fig:2}.

We have incorporated the above free energy (\ref{eq:2-8-0}) into the cell-dynamics 
code~\cite{Gaylord} written by Mathematica TM~\cite{Wolfram}.  We have considered 
the growth of several special forms of circular nucleus which consist of two 
layers of two different phases embedded in another phase to see the possibility 
of the appearance of the long-lived metastable phase.

\begin{figure}[htbp]
\begin{center}
\includegraphics[width=0.7\linewidth]{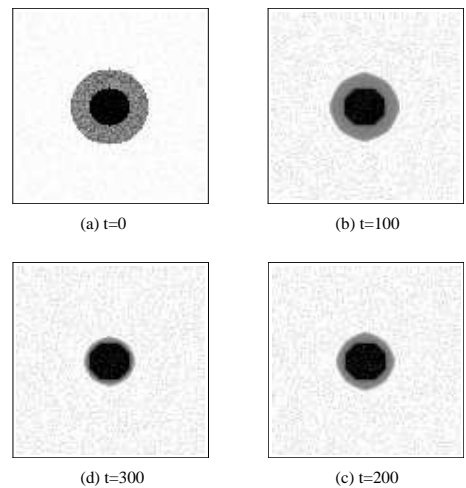}
\caption{
A Gray-level view of the dynamics of a three-layer nucleus when $\epsilon=0.1$. 
The black area is the vapor phase, the gray area is the metastable liquid phase, 
and the white area is the crystal phase.  (a) Initial state is the stable vapor 
nucleus (black) wrapped by the metastable liquid layer (gray) which is further 
embedded in the stable crystal phase (white).  (b), (c) the outer crystal-liquid 
interface shrinks inward by consuming the metastable liquid layer while the 
inner liquid-vapor interface does not expand significantly. (c) Finally the 
stable vapor phase is surrounded by the stable crystal phase and the vapor-crystal 
coexistence is established.}
\label{Fig:8}
\end{center}
\end{figure}

In figure~\ref{Fig:8} we start from the special structure where the stable vapor 
phase is wrapped by the metastable liquid layer, which is further embedded 
in a stable crystal. The initial crystal, liquid and vapor phases are prepared 
by randomly selecting the order parameter $\psi$ from 0.9 to 1.1 for crystal, 
from -0.3 to 0.3 for  liquid and from -1.1 to -0.9 for vapor phases.  From the 
analogy of the expansion of a stable nucleus and the shrinking of metastable 
inner void in a two-phase system, it is expected that the outermost crystal 
phase expands inward and the inner vapor core expands outward by consuming the 
intermediate metastable liquid layer.  Actually, the crystal-liquid interface 
move inward as expected while the liquid-vapor interface does not move 
significantly.  The vapor core does not move significantly even 
if the radius is larger than 
the critical radius as the surrounding metastable liquid layer has finite 
thickness, while the crystal-liquid front shrinks because no critical radius 
exists.  As has been discussed in eq.(\ref{eq:a5}), the capillary pressure
accelerates the shrinking crystal-liquid interface, while it decelerates the 
growing liquid-vapor interface.  Then, the
slow liquid-vapor interface does not have enough time to move because 
metastable liquid is consumed by the fast crystal-liquid interface.  
Finally, the metastable liquid layer disappears completely and the 
stable vapor core is surrounded by the stable crystal phase and the vapor-crystal 
coexistence is established.

\begin{figure}[htbp]
\begin{center}
\includegraphics[width=0.7\linewidth]{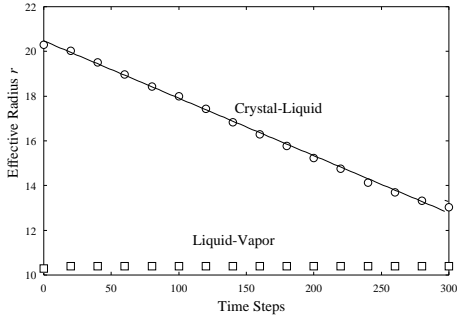}
\caption{
The time evolution of the effective radius of a circular crystal-liquid 
($\bigcirc$) and liquid-vapor ($\square$) interface when $\epsilon=0.1$. The total area is 
100$\times$100=10000.  The radius of the crystal-liquid interface 
decreases while that of the liquid-vapor interface remains almost 
constant.}
\label{Fig:9}
\end{center}
\end{figure}

Figure~\ref{Fig:9} shows the time evolution of the effective radii of the 
crystal-liquid and the liquid-vapor interface estimated from the area of 
the nucleus from (\ref{eq:3-3x}).  We defined the area $S$ of three phases 
as the number of pixels which belong to them; the pixel of vapor is defined 
by the order parameter $\psi<-0.5$, that of crystal by $\psi>0.5$ and that of 
liquid by $-0.5 \ge \psi \ge 0.5$.  Figure~\ref{Fig:9} clearly indicates 
that the effective radius of the crystal-liquid interface decreases with 
constant velocity $v$ while that of the liquid-vapor interface remains 
almost constant.  The crystal-liquid interfacial velocity estimated by 
fitting the straight line to the simulation result is $v=0.026$ which is 
again the same order of magnitude as the interfacial velocity in a two-phase 
system listed in Tab.~\ref{tab:1}.

\begin{figure}[htbp]
\begin{center}
\includegraphics[width=0.7\linewidth]{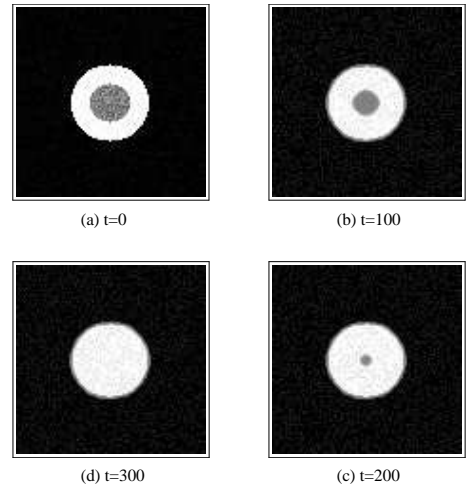}
\caption{
The same as figure~\ref{Fig:8} when $\epsilon=0.1$ but for the different 
ordering of the layers.  (a) Initial state is the metastable liquid void 
(gray) wrapped by the stable crystal (white) layer which is further 
embedded in a stable vapor (black).  (b), (c) the metastable liquid void 
shrinks while the outer stable crystal-vapor interface remains the same.  
(c)Finally the stable crystal phase is surrounded by the stable vapor 
phase and the vapor-crystal coexistence is established.  There is always a
thin layer of liquid at the solid-vapor interface because we use continuous order parameter
$\psi$.}
\label{Fig:10}
\end{center}
\end{figure}

When the metastable liquid core is surrounded by the stable crystal 
and vapor phases, the crystal-liquid interface shrinks and the metastable 
liquid void disappears as shown in Fig.~\ref{Fig:10}.  The vapor-crystal 
interface does not move appreciably because the vapor and the crystal 
phases are in equilibrium and can coexist. Since our cellular dynamics
model uses continuous order parameter $\psi$, the order parameter changes continuously
in space.  Then, there is always a thin layer of liquid with $\psi_{l}=0$ between
the crystal core with $\psi_{c}=1$ and the vapor environment with $\psi_{v}=0$ in Fig.~\ref{Fig:10}.

\begin{figure}[htbp]

\begin{center}
\includegraphics[width=0.7\linewidth]{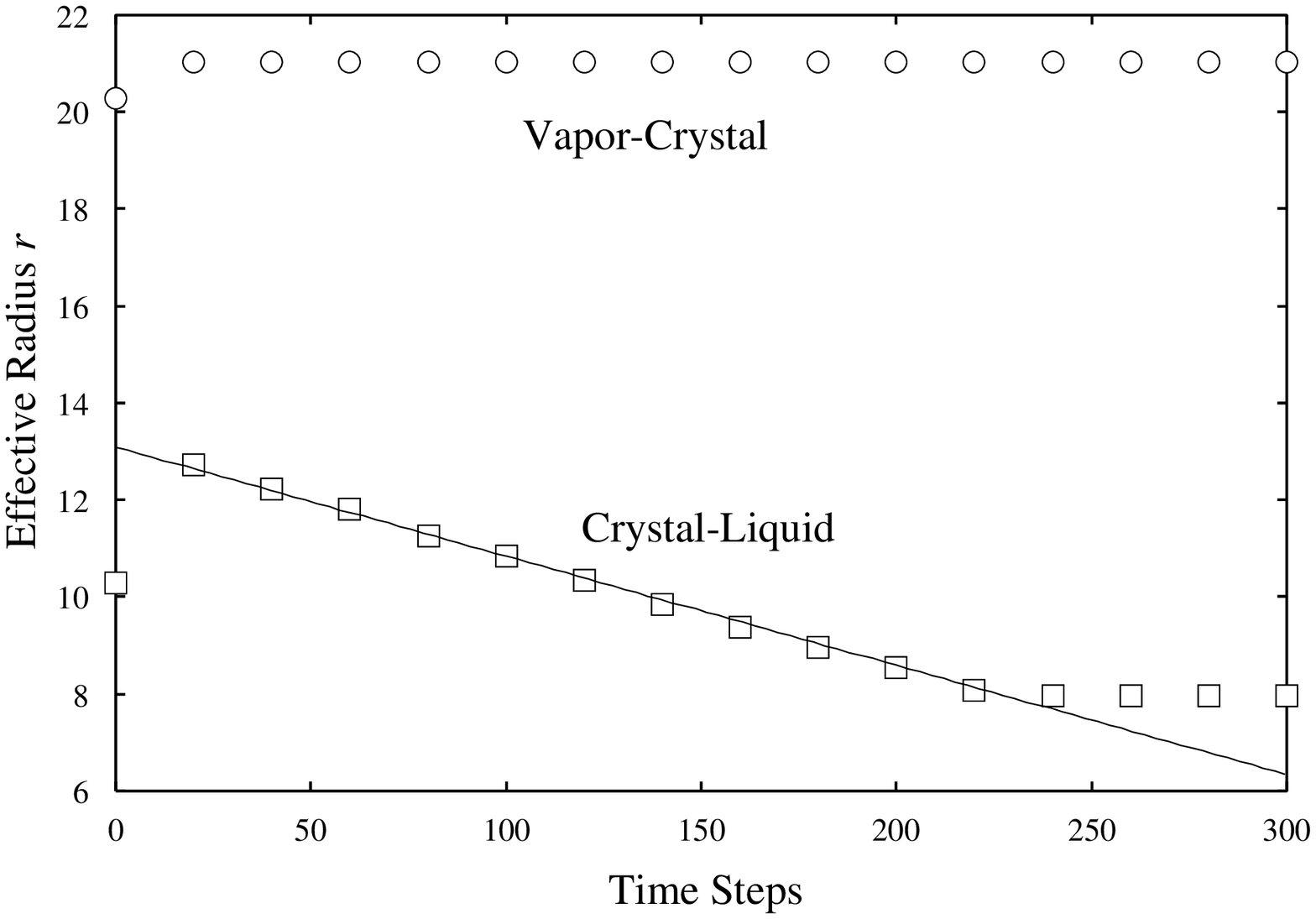}
\caption{
The time evolution of the effective radius of circular vapor-crystal 
($\bigcirc$) and crystal-liquid ($\square$) interface when $\epsilon=0.1$. The radius of 
outer crystal-vapor interface remains the same while that of the 
crystal-liquid interface decreases.}
\label{Fig:11}
\end{center}
\end{figure}

Figure~\ref{Fig:11} shows that the effective radius of the vapor-crystal 
interface remains constant while that of the crystal-liquid interface 
decreases almost linearly. The crystal-liquid interfacial 
velocity is estimated to be $v=0.023$ which is again comparable to 
the values shown in Tab.~\ref{tab:1} for the two-phase system.

These two examples have clearly indicated that the behavior of the 
stable core and the metastable void in a triple-phase system is 
similar to those in a two-phase system.

\begin{figure}[htbp]
\begin{center}
\includegraphics[width=0.7\linewidth]{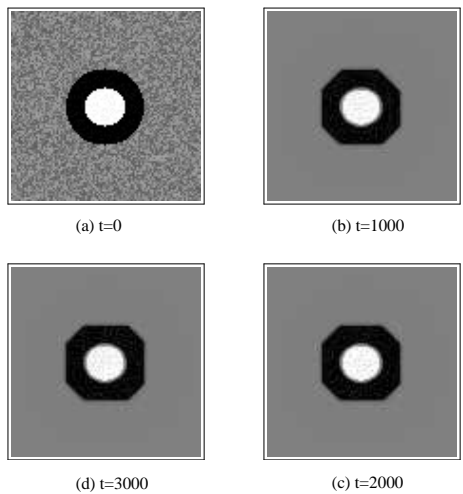}
\caption{
The Gray-level view of the dynamics of a three-layer nucleus when 
$\epsilon=0.1$.  (a) Initial state is the stable crystal core (white) 
wrapped by the stable vapor (black) layer which is further embedded 
in the metastable liquid (gray).  (b), (c), (d) The initial structure 
does not change appreciably even after very long time steps 
$t$=1000, 2000, 3000. Note the difference of time scale 
from Figs.~\ref{Fig:8} and \ref{Fig:10}. Again, there is always a
thin layer of liquid at the solid-vapor interface as in Fig.~\ref{Fig:10}.
 }
\label{Fig:4}
\end{center}
\end{figure}

When the composition of the circular nucleus changes, a very interesting 
behavior is observed.  In Figure~\ref{Fig:4}, we start from the special 
structure where the stable circular crystal core is wrapped by a stable 
vapor layer, which is further embedded in a metastable liquid environment.  
The initial crystal, liquid and vapor phases are prepared by randomly 
selecting the order parameter $\psi$ from 0.9 to 1.1 for  crystal, 
from -0.3 to 0.3 for liquid and from -1.1 to -0.9 for vapor phases 
respectively as before.  Intuitively, we expect that the stable inner 
crystal core may not grow because of the crystal-liquid coexistence 
while the stable outer vapor layer expands outward by consuming the 
metastable liquid environment.  However, we observe that this threefold 
structure is rather stable for a very long time.  The crystal phase 
as well as the vapor phase cannot grow and the metastable liquid phase 
survives and occupies almost the same region for a long time as if 
the vapor-liquid coexistence is locally established at the liquid-vapor 
interface.  

\begin{figure}[htbp]
\begin{center}
\includegraphics[width=0.7\linewidth]{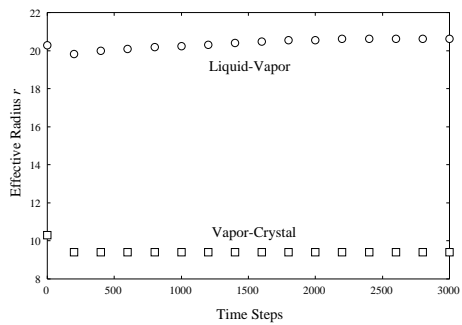}
\caption{
The time evolution of the effective radii of a circular 
vapor-crystal ($\square$) and liquid-vapor ($\bigcirc$) interface
when $\epsilon=0.1$. The total 
area is 100$\times$100=10000.  The two radii which remain almost 
constant indicating that the growth of stable core is prohibited and 
the long-lived liquid phase survives even though it is thermodynamically 
metastable.  Note the long time scale compared with 
Figs.~\ref{Fig:9} and \ref{Fig:11}.   }
\label{Fig:5}
\end{center}
\end{figure}

\begin{figure}[htbp]
\begin{center}
\includegraphics[width=0.7\linewidth]{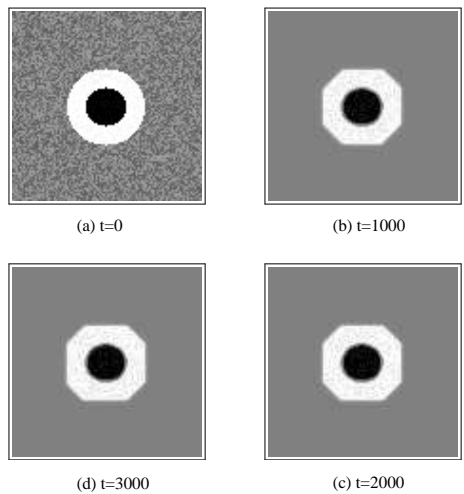}
\caption{
The same as Fig.~\ref{Fig:4}, but now the stable vapor core (black)
is wrapped by the stable crystal (white) layer which is further embedded
in the metastable liquid (gray).  We observe exactly the same morphology
as in Fig.~\ref{Fig:4} if we exchange black (vapor) and white (crystal).
 }
\label{Fig:a1}
\end{center}

\end{figure}

Figure~\ref{Fig:5} shows the time evolution of the effective radius 
of the vapor-crystal and liquid-vapor interface estimated from the 
area of the nucleus calculated from (\ref{eq:3-3x}).  We confirm the 
stable vapor-crystal as well as the stable liquid-vapor interfaces 
which do not move appreciably.  In order to confirm the numerical accuracy of  Fig.~\ref{Fig:4}, we check the evolution of
a dual system where crystal and vapor is exchanged.  Figure \ref{Fig:a1} clearly indicates
that the dual system shows exactly the same behavior as in Fig.~\ref{Fig:4}.

These striking results may be interpreted from the shape of the free 
energy in Fig.~\ref{Fig:1} using the argument of Cahn~\cite{Cahn}.  Since 
the parameter $\epsilon=0.1$, the system remains in the region of phase 
diagram (Fig.~\ref{Fig:2}) where a hidden liquid-vapor binodal exists 
and we can draw a common tangent between the vapor and liquid phases.  
This means that it is possible to establish local liquid-vapor 
coexistence by changing the local pressure or the chemical potential even though the liquid phase is metastable 
(Fig.~\ref{Fig:1} ).  Then, the liquid-vapor interface can not move.  
The faceted structure of the liquid-vapor interface appears probably due to the high symmetry of the problem since we put the nucleus at the center of the area and the nearest and the next-nearest neighbors are used to calculate the Laplacian.  The flat interface is also favorable to mitigate the capillary pressure which acts to expand the liquid-vapor interface.  A similar faceted structure appears also in the stable liquid-vapor interface in Fig.~\ref{Fig:8}.

Furthermore, since this stable vapor layer is so tightly attracted by 
an inner crystal core to maintain vapor-crystal coexistence, the 
vapor-crystal interface also can not move.  Therefore this special 
three-layer structure which Renth {\it et al.}~\cite{Renth} called 
the "boiled-egg crystal" becomes rather stable.  The existence of 
this {\it stable} crystal wrapped by stable vapor layer in metastable 
liquid is predicted theoretically from the shape of the free 
energy~\cite{Cahn,Evans1,Renth} and suggested experimentally in a 
colloid-polymer mixture~\cite{Poon,Poon2}.

We note in passing, that this boiled-egg crystal is in sharp contrast 
to the transient metastable phase predicted from the steady state 
solution of the Ginzburg-Landau equation.  This transient phase appears when 
it is sandwiched by the two stable phase due to the difference of the 
front speed of two interfaces of the stable and metastable 
phase~\cite{Evans1,Bechhoefer,Celestini}, while our long-lived metastable 
phase appears when it surrounds the two stable phases.

\begin{figure}[htbp]
\begin{center}
\includegraphics[width=0.7\linewidth]{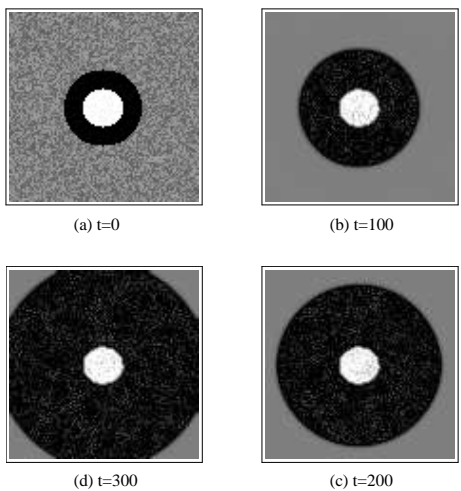}
\caption{
The same as Fig.~\ref{Fig:4} when $\epsilon=0.3$.  (a) Initial state is the stable 
crystal core (white) wrapped by the stable vapor layer (black) which is further 
embedded in the metastable liquid.  (b), (c) The stable vapor layer grows by 
consuming the metastable liquid environment, while the stable crystal core 
remains the same.  (c) Finally the stable crystal phase is surrounded by the 
stable vapor phase and the vapor-crystal coexistence is established.  
}
\label{Fig:6}
\end{center}
\end{figure}

As the parameter $\epsilon$ increases further above 0.2, the hidden liquid-vapor 
coexistence cannot be established (Fig.~\ref{Fig:1}), because we cannot construct 
a common tangent.  Then this boiled-egg crystal structure cannot remain stable 
as shown in Fig.~\ref{Fig:6}.  The vapor phase starts to grow by consuming the 
metastable liquid phase, while core crystal phase remains almost the same size 
and the shape.  Finally the metastable liquid phase disappears and the stable 
crystal phase is wrapped by the stable vapor phases and the vapor-crystal 
coexistence is established.  Here again, the initial crystal, liquid and vapor 
phases are prepared by randomly selecting the order parameter $\psi$ from 
0.9 to 1.1 for crystal, from -0.3 to 0.3 for liquid and from -1.1 to -0.9 for 
vapor phases.

\begin{figure}[htbp]
\begin{center}
\includegraphics[width=0.7\linewidth]{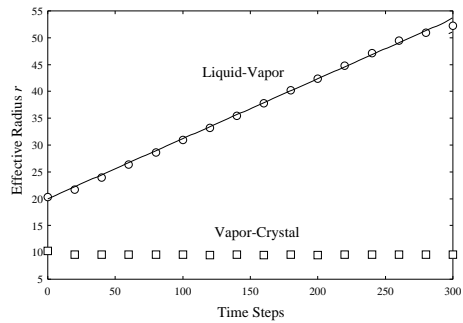}
\caption{
The time evolution of the effective radii of the circular vapor-crystal 
($\square$) and liquid-vapor ($\bigcirc$) interfaces. The total area 
is 100$\times$100=10000.  The radius of liquid-vapor interface increases 
linearly while that of the vapor-crystal interface remains almost constant as 
the function of time.}
\label{Fig:7}
\end{center}
\end{figure}

Figure~\ref{Fig:7} clearly indicates that the effective radius of the 
liquid-vapor interface increases while that of the vapor-crystal interface 
remains the same as the function of time.  The liquid-vapor interfacial 
velocity estimated by fitting the straight line to the simulation data 
is $v=0.112$ which is the same order of magnitude listed in 
Tab.~\ref{tab:1} for the two-phase system.

Similarly, this long-lived boiled-egg structure will be destroyed by the thermal noise,
which can be simulated by using the cell dynamics equation~\cite{Puri}
\begin{equation}
\psi(t+1,n)=F[\psi(t,n)]+B \eta (t,n)
\label{eq:b1}
\end{equation}
instead of (\ref{eq:2-4}), where $B$ is the amplitude of the noise, $\eta(t,n)$ is a uniform
random number between -1.0 and 1.0.

\begin{figure}[htbp]
\begin{center}
\includegraphics[width=0.7\linewidth]{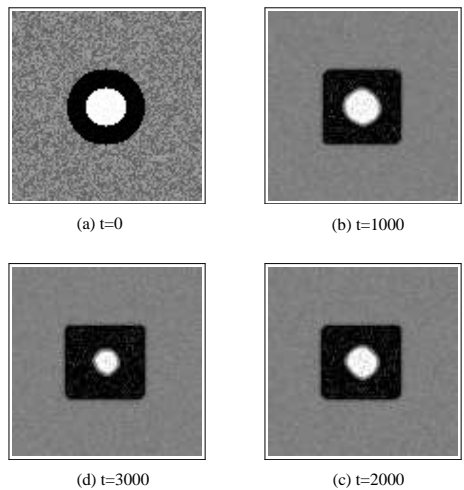}
\caption{
The same as Fig.~\ref{Fig:4} but the thermal noise with
$B=0.07$ is included.  The long-lived metastable state is
destroyed by the thermal noise, however the structure is
still rather stable for a long time.
 }
\label{Fig:a2}
\end{center}
\end{figure}

In Fig.~\ref{Fig:a2}, we start from the same special structure as in Fig.~\ref{Fig:4} when $\epsilon=0.1$ but
the thermal noise with $B=0.07$ is included.  The thermal noise certainly destroys the long-lived metastable
configuration as expected, but it still remains rather stable for a long time.  The thermal noise also
destroys the circular or even the faceted structure and the liquid-vapor interface becomes flat. 
 The thermal noise acts to eliminate the curvature of the interface to suppress the capillary pressure.
  
\begin{figure}[htbp]
\begin{center}
\includegraphics[width=0.7\linewidth]{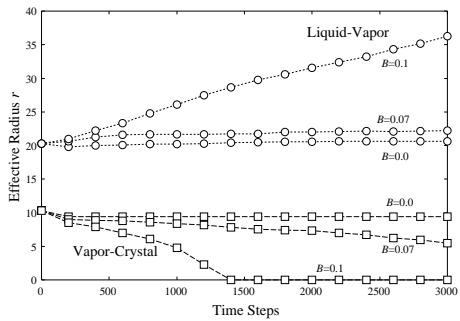}
\caption{
The same as Fig.~\ref{Fig:5} for various noise lever $B=0.00$ (Fig.~\ref{Fig:5}), 
$B=0.07$, and $B=0.1$ when $\epsilon=0.1$.
The effective radius of the liquid-vapor interface starts to increases and that of the
vapor-crystal interface starts to decrease as we increase the noise level $B$. 
However, the time scale is still much longer than Fig.~\ref{Fig:7}.}
\label{Fig:a3}
\end{center}
\end{figure}

Naturally, the larger the thermal noise, the shorter the lifetime of the
metastable configuration as shown in Fig.~\ref{Fig:a3}.  A similar effect of noise
on the time-scale of the evolution was observed
in the TDGL model of nucleation~\cite{Valls} for a non-conserved order
parameter.

These last four examples indicate that the kinetics of phase transformation 
is definitely affected by the presence of metastable phase and the hidden 
binodals which is in no way related to the equilibrium phase diagrams. The 
long-lived metastable phase could appear macroscopically if it accommodates 
the special composite nucleus which consists of a stable crystal core 
surrounded by an equally stable vapor layer.  

\section{Conclusion}
\label{sec:sec4}

In this paper, we have used the cell-dynamics method to study the evolution 
of a single composite nucleus.  We have studied the three-phase system which 
has a hidden binodal with two stable and one metastable phases.  We have 
called two stable phases, crystal and vapor, and one metastable phase liquid. 
We could successfully simulate the evolution of stable phases and the 
regression of metastable phase. We have found, however, one special 
configuration of a stable crystal core wrapped by a stable vapor layer 
embedded in the metastable liquid environment becomes stable and 
stationary for long time.  This means that the long-lived metastable 
environment phase can persist and can appear as a macroscopic phase even 
though it is thermodynamically metastable.  According to the argument 
of Cahn~\cite{Cahn}, this result can be interpreted from the hidden 
liquid-vapor binodal which can be constructed from the common tangent between 
the stable vapor and metastable liquid phases.

In conclusion, we have used a cell dynamics method to study the growth 
of a single nucleus which has traditionally been explained using the partial 
differential equation derived from time-dependent-Ginzburg-Landau 
equation~\cite{Valls} or the so-called phase field model~\cite{Castro,Granasy2}.  
We have found that the long-lived metastable phase can appear during the 
phase transformation as predicted by several researchers~\cite{Ostwald,Cahn,Poon,Renth}.  
This cell-dynamics method is not only flexible but numerically stable to 
handle such a complex situation when many phases can coexist.  Further 
extension and modification of the method to explain the formation of 
various intermediate phases will be possible.

\begin{acknowledgments}
The author is grateful to Dr. M. Nakamura for enlightening
discussion.  This work is partially supported by the grant
from the Ministry of Education, Sports and culture of Japan.
\end{acknowledgments}

\newpage 
\bibliography{myfile3x}

\end{document}